\documentstyle[epsfig]{l-aa}

\begin{document}

\thesaurus{06(08.05.2;08.02.1;08.14.1;13.25.5)}

\title{On the nature of Be/X-ray binaries}

\author{Ignacio~Negueruela}                   
                                                            
\institute{Astrophysics Research Institute, Liverpool John Moores University, 
Byrom St., Liverpool, L3 3AF}

\offprints{ind@astro.livjm.ac.uk}

\date{Received April 1 ; accepted June 23, 1998}

\maketitle                        
\markboth{Negueruela: The nature of Be/X-ray binaries}{Negueruela:
The nature of Be/X-ray binaries}

\begin{abstract}
It has been suggested that most Be/X-ray binaries are low X-ray luminosity 
nearby objects, containing white dwarfs (Chevalier \& Ilovaisky 1998). 
We show that existing evidence indicates that all known Be/X-ray binaries 
are relatively bright X-ray sources containing neutron stars and that 
the spectral distribution of this group differs considerably from that of 
isolated Be stars. We suggest that the different 
X-ray properties of the systems can be explained by the sizes of the orbits of 
the neutron stars. Systems with close orbits are bright transients which show 
no quiescent emission as a consequence of centrifugal inhibition of accretion. 
Systems with wide orbits are persistent sources and display no large 
outbursts. Systems with intermediate orbits present a mixture of both 
behaviours.

\end{abstract}

\keywords{stars: emission line, Be --  binaries: close -- neutron   
-- X-ray: stars}

\section{Introduction}
Be/X-ray binaries are X-ray sources composed of a Be star and a compact object.
The high-energy radiation is believed to arise due to accretion of material 
associated with the Be star by the compact object. 

The name ``Be star'' 
is used as a general term describing an early-type non-supergiant star, which 
at some time has shown emission in the Balmer series 
lines (Slettebak 1988, for a review). Both the emission
lines and the characteristic strong infrared excess when compared to
normal stars of the same spectral types are attributed to the presence
of circumstellar material in a disc-like geometry. The causes that give rise
to the disc are not well understood. Different mechanisms (fast rotation, 
non-radial pulsation, magnetic loops) have been proposed, but it seems that
none of them can explain the observed phenomenology on its own. The discs
are rotationally dominated (Hanuschik 1996), but UV spectra of Be stars show 
evidence of a high-velocity low-density wind, suggesting that
mass-loss from Be stars takes the shape of a fast radiative wind in the polar 
regions and a slow higher-density outflow in the equatorial regions, which
generates the disc (Lamers \& Waters 1987). It is generally believed that
the material forming the disc accelerates radially at distances larger than
those probed by the optical emission lines (see Chen \& Marlborough 1994, 
Okazaki 1997). X-ray activity in Be/X-ray binaries would then be due to the 
interaction
of the neutron star with this radial outflow (Waters et al. 1988)

Be/X-ray binaries can present 
very different states of X-ray activity (Stella et al. 1986):
\begin{itemize}
\item Persistent low-luminosity ($L_{{\rm x}} \la 
10^{36}$ erg s$^{-1}$) X-ray emission or no detectable emission. 
\item Short (a few days) X-ray outbursts ($L_{{\rm x}} 
\approx 10^{36} - 10^{37}$ erg s$^{-1}$) separated by the orbital period (Type 
I outbursts), generally (but not always) occurring close to the time of 
periastron passage of the neutron star.
\item Giant (Type II) X-ray outbursts ($L_{{\rm x}} 
\ga 10^{37}$ erg s$^{-1}$), which do not show clear orbital modulation and last
several weeks. 
\end{itemize}

\begin{table*}[ht]
\caption{Known galactic Be/X-ray binaries with detected X-ray pulsation and 
their basic
parameters. Orbital periods marked with `$^{*}$' represent the recurrence time 
of X-ray outbursts and not orbital solutions. Objects for which the orbital 
period is noted as `large' are persistent low-luminosity X-ray sources, likely
to have periods of a few hundred days.
Spectral types marked `$^{*}$' are estimated from photometry and the distances 
derived should be treated with caution.
Objects for which no quiescence luminosity is given have been detected only 
during outbursts. The distance to EXO\,2030+375 and its luminosity,
estimated from the change rates in spin period and X-ray luminosity, are from
Parmar et al. (1989).}
\begin{center}
\begin{tabular}{lccccccc}
\hline
Name & $P_{{\rm s}}$(s) & $P_{{\rm orb}}$(d) & Optical & Spectral & Distance & Quiescence $L_{{\rm x}}$ & Maximum $L_{{\rm x}}$\\
& & & Counterpart & Type & & (erg s$^{-1}$) &  (erg s$^{-1}$)\\
\hline
&&&&&&&\\
4U\,0115+634 &3.6 & 24.3 &V635 Cas & B0.2V$^{\mathrm{a}}$ & 6 kpc & $\sim 10^{36}$$^{\mathrm{b}}$& $\sim 10^{38}$$^{\mathrm{b}}$\\
RX\,J0146.9+6121 & 1412 & large & LS\,I$\;$+61$^{\circ}$235 & B1V$^{\mathrm{c}}$ & 2.3 kpc & $2-4\times10^{34}$$^{\mathrm{d}}$ & $\sim 5\times10^{35}$$^{\mathrm{d}}$\\
V\,0332+53 & 4.4 & 34.2 & BQ Cam & O8.5V$^{\mathrm{a}}$ & 7 kpc & $-$& $\sim 2\times10^{38}$$^{\mathrm{e}}$\\
4U\,0352+30 & 837& large& X Per & B0V$^{\mathrm{f}}$ & 700 pc & $2\times10^{34}$$^{\mathrm{d}}$ & $\sim 10^{35}$$^{\mathrm{d}}$\\
A\,0535+262 & 103 & 111 & V725 Tau & B0III$^{\mathrm{g}}$ & 2.0 kpc &$\sim 2\times10^{35}$$^{\mathrm{h}}$ & $\sim 2\times10^{37}$$^{\mathrm{h}}$ \\
A\,0726$-$26 & 103.2 & 35$^{*}$ & LS 437 & O8.5V$^{\mathrm{i}}$ & 6 kpc &$\sim 3\times10^{35}$$^{\mathrm{j}}$ &$\sim 10^{36}$$^{\mathrm{k}}$\\
GRO J1008$-$57 &  93.5  & 248$^{*}$ & star & B0$^{*}$$^{\mathrm{l}}$& $\sim$ 5kpc &$\sim5\times10^{34}$$^{\mathrm{m}}$&$\sim2\times10^{37}$$^{\mathrm{n}}$\\
A\,1118$-$616 & 406.5 & large & Wray 977 &O9.5V$^{\mathrm{o}}$& 6 kpc& $\sim 3\times10^{34}$$^{\mathrm{o}}$&$\sim 5\times10^{36}$$^{\mathrm{p}}$\\
4U\,1145$-$619 & 292 & 188$^{*}$ & V801 Cen & B0.7V$^{\mathrm{q}}$ & 3.1 kpc &$\sim 10^{35}$$^{\mathrm{r}}$ & $\la 10^{37}$$^{\mathrm{q}}$\\
4U\,1258$-$61 & 272 &132.5$^{*}$ & V850 Cen & B2V$^{\mathrm{s}}$ & 2.5 kpc & $\sim10^{35}$$^{\mathrm{s}}$ & $\sim10^{36}$$^{\mathrm{s}}$  \\
2S\,1417$-$624 & 17.6 & 42.1 & star &B1V$^{\mathrm{t}}$ & 6 kpc & $-$ &$\ga 10^{37}$$^{\mathrm{u}}$\\
EXO\,2030+375 & 41.7 &46 & star & B0$^{*}$$^{\mathrm{v}}$ & 5.3 kpc & $-$ & $\sim 10^{38}$$^{\mathrm{w}}$\\
Cep X$-$4 & 66.3 & ? & star & B1$^{*}$$^{\mathrm{x,y}}$ & 4 kpc & $\sim 5\times10^{33}$$^{\mathrm{x}}$ & $\sim 10^{37}$$^{\mathrm{x}}$\\
\hline
\hline
\end{tabular}
\begin{tabbing}
circus\=$^{\mathrm{a}}$ Negueruela et al., in prepar\=$^{\mathrm{a}}$ Negueruela et al., in prep.\=
$^{\mathrm{a}}$ Janot-Pacheco et al. (1981)\= \kill
\>$^{\mathrm{a}}$ Negueruela et al., in prep.\> $^{\mathrm{b}}$ Campana (1996)\>
$^{\mathrm{c}}$ Reig et al. (1997a)\> $^{\mathrm{d}}$ Haberl et al. (1998)\\
\>$^{\mathrm{e}}$ Whitlock (1989)\> $^{\mathrm{f}}$ Lyubimkov et al. (1997)\>
$^{\mathrm{g}}$ Steele et al. (1998)\> $^{\mathrm{h}}$ Sembay et al. (1990)\\
\>$^{\mathrm{i}}$ Negueruela et al. (1996)\> $^{\mathrm{j}}$ Corbet \& Peele (1997)\>
$^{\mathrm{k}}$ Steiner et al. (1984)\> $^{\mathrm{l}}$ Coe et al. (1994)\\
\>$^{\mathrm{m}}$ Macomb et al. (1994)\> $^{\mathrm{n}}$ Wilson et al. (1994)\>
$^{\mathrm{o}}$ Janot-Pacheco et al. (1981)\> $^{\mathrm{p}}$ Motch et al. (1988)\\
\>$^{\mathrm{q}}$ Stevens et al., in prep.\> $^{\mathrm{r}}$ White et al. (1980)\>
$^{\mathrm{s}}$ Parkes et al. (1980)\> $^{\mathrm{t}}$ Grindlay et al. (1984)\\
\>$^{\mathrm{u}}$ Finger et al. (1996)\> $^{\mathrm{v}}$ Coe et al. (1988)\>
$^{\mathrm{w}}$ Parmar et al. (1989)\> $^{\mathrm{x}}$ Bonnet-Bidaud \& Mouchet (1998)\\
\>$^{\mathrm{y}}$ Roche, priv. comm.\>
\end{tabbing}
\end{center}
\label{tab:galax}
\end{table*}

Some systems only display persistent emission, but most of them show outbursts 
and are termed Be/X-ray transients. Both kinds of systems seem to fall in a
relatively narrow region of the $P_{{\rm orb}}/P_{{\rm spin}}$ diagram, 
known as Corbet's diagram (Corbet 1986; see also Waters \& van Kerkwijk 1989).

Based on distance measurements to several proposed counterparts of Be/X-ray 
binaries by the {\em Hipparcos} satellite, Chevalier \& Ilovaisky (1998, 
henceforth CI98) have suggested that the compact object in most Be/X-ray 
binaries is a white dwarf (WD) and that the class of objects can be 
characterised
as nearby low-luminosity sources. In this paper, we set out to show that the
existing evidence does not favour that interpretation, and that Be/X-ray
binaries contain mostly neutron stars.

\section{The sample of Be/X-ray binaries}

CI98 use a sample of 13 proposed counterparts to Be/X-ray binaries. Their 
sample is limited to objects with $V \la 12$ so that they can be observed 
with {\em Hipparcos}. Seven of their sources are unconfirmed candidates to
faint unidentified hard X-ray sources 
observed during the $HEAO-1$ all-sky survey with the Modulation Collimator.
Tuohy et al. (1988) proposed their association with Be 
stars on the basis of positional coincidence.  Because of the large error 
boxes, Tuohy et al. (1988) warned that several of these identifications 
could be spurious.
Since no further detection of any of these sources has been reported, the
question of their identification and the real nature of these X-ray 
sources remains open. This has not been taken 
into account by CI98. Moreover, their sample is magnitude-limited 
and necessarily includes only nearby sources (since there is 
only a limited range of absolute magnitudes for Be stars),

In order to compare these candidates with more secure identifications
of Be/X-ray binaries, we set out to select a more appropriate sample. 
In Table \ref{tab:galax}, we have listed known 
galactic Be/X-ray binaries with detected X-ray pulsation and a 
proposed optical counterpart. Hard X-ray spectra
and pulsations are the most typical characteristics of a Massive X-ray 
Binary. Distances in Table \ref{tab:galax} are derived from the spectral 
type of the counterpart, assuming that they have the average optical 
luminosity for their spectral type, as given by Vacca et al. (1996) or 
Schmidt-Kaler (1982) -- except for EXO\,2030+375 (see caption to Table 
\ref{tab:galax}). X-ray luminosities have been calculated using 
these distances. No attempt has been made to take 
into account errors due to the uncertainty in the spectral classification or 
in the luminosity corresponding to a given spectral type, since they are not 
supposed to be systematic. An important point to be considered here, 
relevant for the following discussion, is that the optical counterparts to
Be/X-ray binaries are supposed to have the same physical characteristics as
normal Be stars of the same spectral type.. Detailed simulations by 
Vanbeveren \& de Loore (1994) and 
de Loore \& Vanbeveren (1995), in which Be/X-ray binaries are formed from 
moderately massive close binaries that undergo mass transfer, show that
the properties of the Be star are those of a normal star of the same mass,
at least while it remains in the main sequence. Under certain circumstances, 
the star can become an overluminous supergiant at a later stage.

Table \ref{tab:lmcex} lists all known Be/X-ray binaries in the 
Magellanic 
Clouds (MCs) with detected X-ray pulsation and a proposed optical counterpart. 
X-ray luminosities in this table, taken from the literature, are calculated 
assuming standard distances to the MCs.

\begin{table}[ht]
\caption{Known Be/X-ray binaries in the Magellanic Clouds with detected X-ray 
pulsation and their basic parameters. The only source for which the orbital 
period is known is A\,0535$-$668, with a 16.7-d period derived from the 
recurrence time of X-ray outbursts.}
\begin{center}
\begin{tabular}{lccc}
\hline
Name & $P_{{\rm s}}$(s) & Spectral & Max $L_{{\rm x}}$\\
& & Type & (erg s$^{-1}$)\\
\hline
&&&\\
2E\,0050.1$-$7247 & 8.9 & ? & $\sim 10^{36}$$^{\mathrm{a}}$\\
1WGA\,J0053.8$-$7226 & 41.6$^{\mathrm{b}}$ & B1V$^{*}$$^{\mathrm{c}}$& ? \\
A\,0535$-$668 & 0.07 &B2IV$^{\mathrm{d}}$ & $\sim 10^{39}$$^{\mathrm{d}}$ \\
RX\,J0502.9$-$6626 & 4.1 & B0III$^{\mathrm{e}}$ & $\sim 4\times10^{37}$$^{\mathrm{f}}$\\
EXO\,0531.1$-$6609 & 13.6 & ?& $\sim 1\times10^{37}$$^{\mathrm{g}}$\\
RX\,J0529.8$-$6556 & 69.5&$\sim$ B2V$^{\mathrm{h}}$ &$\sim 10^{36}$$^{\mathrm{h}}$\\
\hline
\hline
\end{tabular}
\end{center}
\begin{tabbing}
cir\=$^{\mathrm{a}}$ Negueruela et al., in prepar\=$^{\mathrm{a}}$ Negueruela et al., in prep.\=\kill
\>[$^{\mathrm{a}}$] Israel et al. (1997)\>
[$^{\mathrm{b}}$] Corbet et al. (1998)\\
\>[$^{\mathrm{c}}$] Buckley et al., in prep.\>
[$^{\mathrm{d}}$] Charles et al. (1983)\\
\>[$^{\mathrm{e}}$] Crampton et al. (1985)\>
[$^{\mathrm{f}}$] Schmidtke et al. (1995)\\
\>[$^{\mathrm{g}}$] Burderi et al. (1998)\>
[$^{\mathrm{h}}$] Haberl et al. (1997)\\
\end{tabbing}
 \label{tab:lmcex}
\end{table}


\section{Spectral distribution}

As can be seen in Tables \ref{tab:galax} and \ref{tab:lmcex}, all the 
optical counterparts to galactic and MC sources have spectral types 
earlier than B2, and there are several Oe stars. Most objects 
have firm spectroscopic classifications. A few have spectral 
classifications based on photometric colours or continuum fitting 
and, due to the intrinsic reddening of 
Be stars, could be slightly earlier than classified. Also within this
spectral range are the optical counterparts to Be/X-ray binaries with
no detected pulsations -- LS\,I$\;$+61$^{\circ}$303 (B0V, Steele et al. 
1998), BD\,$+53^{\circ}$2790 (O9.5III, Hiltner \& Bautz 1963), 
RX\,J0117.6$-$7330 ($\sim$B1III, Coe et al. 1998) -- and 
all the probable counterparts to likely Be/X-ray binaries in the 
Magellanic Clouds proposed by Crampton et al. (1985) and Schmidtke et al.
(1994) -- e.g., RX\,J0501.6$-$7034, RX\,J0520.5$-$6932.

\begin{figure}[ht]
\begin{center}
    \leavevmode
\epsfig{file=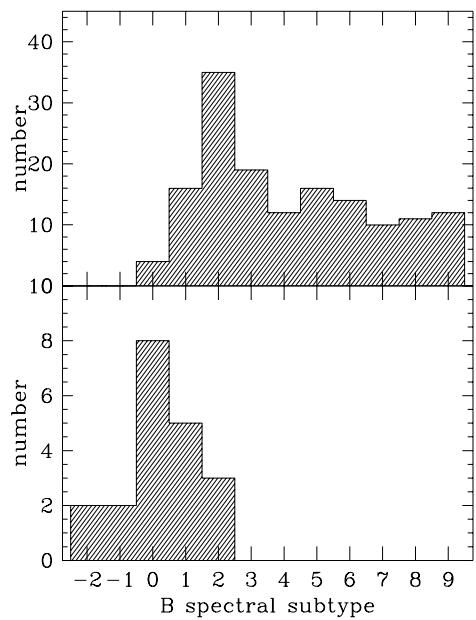, width=6.4cm, bbllx=145pt,
 bblly=440pt, bburx=315pt, bbury=660pt, clip=}
 \end{center}
\caption{The spectral distribution of isolated Be stars is compared to 
that of Be/X-ray binary optical components. Negative spectral subtypes
are used to represent O-type stars. {\bf Top panel:} The spectral
distribution of 150 Be stars present in the BSC, after Porter (1996).
{\bf Bottom panel: }The spectral distribution of 20 Be/X-ray binary 
components, comprising 13 pulsars with spectroscopic spectral-type
determinations, 4 pulsars with photometric spectral-type estimation and
3 sources without detected pulsations with spectroscopic determinations.}
\label{fig:dist}
\end{figure}

The distribution of isolated Be stars 
is completely different. The number of Oe stars is very low, but the 
distribution rises sharply at B0, peaking around B2 and then falls down 
gradually extending up to at least 
spectral type A0 (Slettebak 1988). In Fig. \ref{fig:dist}, the spectral 
distribution of optical components of Be/X-ray binaries is compared with 
a sample of 150 bright Be stars
taken from the catalogue of Slettebak (1982), after Porter (1996). A 
Kolmogorov-Smirnov test of the probability that both samples are extracted
from the same population gives a K-S statistic $D=0.84$ with a significance
of $5.3\times10^{-12}$, clearly indicating that the two samples are extracted
from different populations (a $\chi^{2}$-test gives a reduced $\chi^{2}$ of 
7.1).  

In order to assess the statistical significance of this result, we must
considerer the possible biases in the selection of the two samples
compared. The Be star list contains the majority of Be stars in the
Bright Star Catalogue (BSC) and it is therefore limited by their optical 
magnitude. The BSC contains stars brighter than $V\la6.5$ and it is 
therefore biased towards earlier spectral types. As a consequence, in a 
volume-limited sample, the peak of the distribution would be towards later
spectral types. Abt (1987) found the maximum of the distribution to be at
B3-B4 for a volume-limited sample of field Be stars. In the BSC sample, 
the higher proportion of Be stars in comparison with normal B stars (27\%)
occurs at B4 (Jaschek \& Jaschek 1983). 

The sample of Be/X-ray binaries, on the other hand, is limited by their 
X-ray luminosity. The spectral distribution of this sample could be biased
if there exists a direct correlation between spectral type of the optical
component and the X-ray luminosity, i.e., if there are Be/X-ray binaries
containing late-type Be stars, but all of them are very weak X-ray sources.
However, there are two strong arguments against this hypothesis. First, 
there is no evidence of any dependence of the X-ray luminosity with
spectral type among the known Be/X-ray binaries -- including those in the 
Large Magellanic Cloud, which are all at approximately 
the same distance. The bright transients approaching Eddington luminosity 
extend over the whole spectral range with the brightest transient known 
(A\,0535$-$668) having the latest spectral type (B2IV). This is in clear
contrast with the sharp cut-off at B3.
Second, there is no known correlation 
between the observable properties of Be stars and their spectral type. 
The sizes of their envelopes (as reflected in the emission lines) do not 
seem to depend at all on spectral type. However, if a correlation was to
exist between spectral type and X-ray luminosity, it would imply that there
is a fundamental difference in the mass-loss processes taking place in
early-type and late-type Be stars.

From the above arguments, we conclude that the difference 
seen between the spectral distributions of field 
Be stars and optical components of Be/X-ray binaries must reflect a real
difference in the populations from which they are drawn. 
With a sample of 20 objects all earlier than B3, it seems unlikely 
that any optical member of a Be/X-ray binary is going to have a later 
spectral type.

The early limit in the spectral range of Be/X-ray binary components could 
be simply due to the cessation
of the Be phenomenon at earlier spectral types. Although a few O7e stars are 
known (Conti \& Leep 1974), they are very rare.
The upper limit is in broad agreement with the predictions of the 
models of close binary evolution by Van Bever \& Vanberen (1997). Models in 
which a large amount of angular momentum per unit mass is lost from the 
system during non-conservative mass transfer predict no Be + neutron star 
binaries with 
late-type Be stars (Portegies Zwart 1995; Van Bever \& Vanberen 1997).
The distribution shown in Fig. \ref{fig:dist} indicates that all the 
Be stars with neutron star companions have masses $M_{*}\ga 8-9M_{\sun}$.

\section{A phenomenological model}

The X-ray characteristics of the confirmed Be/X-ray candidates are 
sufficiently consistent to derive a phenomenological model for these systems. 
The low-luminosity persistent X-ray emission seen in many objects 
is due to accretion of low-density material.
This could be the fast polar wind, but it is more likely to be the equatorial 
outflow beyond the regions in which motion is rotationally dominated (and 
where the optical emission lines form), since X-ray emission from 
4U\,1258$-$61 ceased completely when the disc around the companion star 
disappeared even though the polar wind should still be present (Corbet et 
al. 1986).  In the sources with short pulsation (and therefore 
orbital) periods, this quiescent emission is prevented by centrifugal 
inhibition of accretion (Stella et al. 1986): due to the fast rotation and 
strong magnetic field of the neutron star, matter
approaching the magnetosphere is shocked by supersonic rotation and ejected 
beyond the accretion radius (propeller mechanism). 

Previous authors (e.g., Corbet 1986) have assumed that the long periods 
without
outbursts are due to the shrinkage of the disc and that series of outbursts
take place after discrete episodes of mass ejection from the Be star. The
results of Reig et al. (1997b) point very strongly to the possibility that 
the size of the disc is limited by the orbit of the neutron star, presumably 
due to tidal truncation (Okazaki 1998). The existence of X-ray outbursts, 
indicating that the neutron star interacts with material from
the dense regions of the disc, implies that the density distribution in 
the disc can differ from this quiescence configuration. Negueruela et al. 
(1998) have shown how the presence of a density wave in the disc can provide
such a perturbed configuration. Systems with small orbits will then accrete
from very dense regions and become high-luminosity transients. Systems with
wider orbits, in which centrifugal inhibition does not occur, accrete from
less dense regions and show smaller outbursts. Like the transients, 
A\,0535+262 
and GRO\,J1008$-$57 display both Type I and Type II outbursts, but in the
case of 4U\,1145$-$619 and A\,1118$-$616 the distinction is not so clear.
In systems with relatively wide orbits, outbursts can only occur close to 
periastron passage (e.g., A\,0535+262), but in closer systems
they can take place at different orbital phases, depending on the actual 
density distribution in the disc, e.g.,  recent outbursts at phase 
$\sim 0.3$ from 4U\,0115+634 (Negueruela et al. 1998) and 
at phase $\sim 0.5$ from 2S\,1417$-$624 (Finger et al. 1996).

The two systems with longer spin periods, X Per and LS\,I$\;$+61$^{\circ}$235 
have never been observed to undergo X-ray outbursts. In both cases, however, 
long periods of increased X-ray luminosity have been observed (see Haberl 
et al. 1998). Given the known relationship between the spin and orbital 
periods of Be/X-ray binaries (Corbet 1986), both systems are expected to 
have very long orbital periods (many hundred days). The neutron star in a 
very wide orbit can only accrete from the outer low-density regions
of the circumstellar disc, which explains the low X-ray luminosity.
The long periods of increased X-ray luminosity seen in both systems could 
be associated with the dissipation of the discs (Haberl et al. 1998). Type II 
outbursts are very likely to be the corresponding events in systems with 
close orbits.

\section{Discussion}

The X-ray luminosities of all the objects listed in Tables \ref{tab:galax}
and \ref{tab:lmcex} are too high for the expected 
luminosities of Be + WD binaries, estimated to be in the range 
$10^{29}-10^{33}$ erg s$^{-1}$ (Waters et al. 1989), indicating that they
contain neutron stars. It is worth noting that only three of these sources 
could be observed by $Hipparcos$. The distances given in 
Table \ref{tab:galax} for A\,0535+262 and 4U\,1145$-$619 are those 
derived from their spectral types, since Steele et al. (1998) have shown 
that the distances to A\,0535+262 and LS\,I$\;$+61$^{\circ}$303 calculated
by {\em Hipparcos} (which are based on very poor astrometric solutions) 
are inconsistent with several other 
distance indicators (and, at least in the case of A\,0535+262, its X-ray 
spectrum, which can only be explained
in terms of accretion on to a neutron star). This could also be true of the 
distance to 4U\,1145$-$619. Since $\gamma$ Cas seems unlikely 
to be a binary X-ray source (Smith 1997), the sample of objects
with accurate distances in CI98 consists of only one confirmed Be/X-ray 
binary and eight unconfirmed identifications. Six of these objects have 
spectral types later than B3 (up to B8) and therefore there is an almost
negligible statistical probability
that they are extracted from the same population as the optical components 
of standard Be/X-ray binaries. Moreover,  if these identifications are 
correct, they represent a class of objects with much lower X-ray 
luminosities than those of our sample of Be/X-ray binaries (and this 
also applies to the two objects with spectral types in the acceptable range, 
HD 34921 and BZ Cru). The conclusion is that, if the identifications are
correct, they represent a class of objects extracted from a different 
population to the standard Be/X-ray binaries.

Can they represent a sample of the population of Be + WD binaries?
Since we have no previous sample of this population, we do not know its
spectral distribution. Indeed, the only known white dwarf orbiting a 
massive star is the companion of the B5V star HR2875 (Vennes et al. 1997). 
However, there is a major drawback to this interpretation: if
the X-ray activity of these sources is attributed to accretion on to a white 
dwarf, centrifugal inhibition is not a possibility and there is no reason 
why these systems should not be persistent X-ray sources. However, none of
these sources has been detected during the Rosat All-Sky Monitor survey, in
spite of thorough searches for possible binaries (Meurs et al. 1992; 
Motch et al. 1997). Bergh\"{o}fer et al. (1996) list BZ Cru, $\mu^{2}$ Cru
and HD109857 (the three objects in the sample that appear in the BSC) as 
non-detections. No detections of any of the objects have been reported since 
the discovery paper, where it is reported that BZ Cru and HD 34921 (the
two B0 counterparts) had been observed by other satellites 
(Tuohy et al. 1988).

If the identification of these X-ray sources with the proposed
Be stars is real, they represent a population of very low luminosity 
transients. These low-luminosity transients cannot be explained in terms
of Be + WD binaries or neutron stars in very wide orbits -- since these
objects would not be transients. The simplest explanation is that most of 
these counterparts -- if not all -- are really field Be 
stars and not accreting binaries, i.e., they are not optical counterparts
to X-ray sources. This is not surprising, given that they were proposed
only because of positional coincidence with very large error boxes. The 
sample used by CI98 is, in consequence, not representative of Be/X-ray
binaries and therefore their conclusions do not apply to these systems.

\section{Conclusions}

We have studied the global characteristics of Be/X-ray binaries by comparing
the properties of different systems. We find that all the optical counterparts
have spectral types in the range O8-B2, which represents a distribution very
different from that of isolated Be stars. The very different spectral type
distribution of Be/X-ray binaries and isolated Be stars sets strong limits
on acceptable models of close binary evolution.

We have developed a coherent model to explain the different 
X-ray properties of Be/X-ray binaries, in which the main parameter is the 
size of the orbit of the neutron stars. Systems with 
close orbits are fast spinners and show no quiescence emission as a 
consequence of centrifugal inhibition. When the density distribution in the
circumstellar disc of the Be star becomes very asymmetric, they become
bright transients. Systems with wide orbits are persistent sources accreting
from a low-density radial outflow and display no large outbursts. 
Systems with intermediate orbits present a mixture of both behaviours.

We have shown that the sample recently used to conclude that Be/X-ray binaries
are low-luminosity X-ray sources containing white dwarfs consists of objects
extracted from a different population and therefore it is not relevant to
the study of Be/X-ray binaries.

\section*{Acknowledgements}

I would like to thank James B. Steves and Paul Roche for communicating
results in advance of publication and to Mark Finger for communicating his
results on 2S\,1845$-$024. I have benefited from many relevant discussions
with Pablo Reig and Simon Clark

I.~A.~Steele and J.~M.~Porter are thanked for their careful reading of 
the first draft of this paper and their constructive criticism. Special 
thanks to
John Porter for providing the author with the data on isolated Be stars. The
author is supported by a PPARC postdoctoral research assistantship.

\end{document}